\documentstyle[aps,prl,preprint,floats,epsfig]{revtex}

\begin{document}

\tighten

\preprint{\tighten\vbox{\hbox{\bf CLNS 98/1572}
                        \hbox{\bf CLEO 98-9}
                        \hbox{\date{\today}}
}}

\title{\boldmath First Observation of $\Upsilon$(1S)$\to\gamma\pi\pi$ }

\author{CLEO Collaboration}
\date{\today}

\maketitle
\tighten

\begin{abstract}
We report on a study of exclusive radiative decays of the
$\Upsilon$(1S) resonance 
%in a sample of 1.86 million $\Upsilon$(1S) decays 
collected with the CLEO~II detector operating at CESR. 
We present the first observation of the radiative decays 
$\Upsilon$(1S)$\to\gamma\pi^+\pi^-$ and $\Upsilon$(1S)$\to\gamma\pi^0\pi^0$.
For the dipion mass regime $m_{\pi\pi}>$1.0 GeV
we obtain
${\cal B}(\Upsilon$(1S)$\to\gamma\pi^+\pi^-)$=$(6.3\pm 1.2\pm 1.3)
\times 10^{-5}$ and
${\cal B}(\Upsilon$(1S)$\to\gamma\pi^0\pi^0)$=$(1.7\pm 0.6\pm 0.3)
\times 10^{-5}$.
The observed $\gamma\pi\pi$ events
%over background in the dipion mass region 1.2-1.4~GeV 
are consistent 
with the hypothesis
$\Upsilon$(1S)$\to\gamma f_2$(1270). 
\end{abstract}


\newpage

\renewcommand{\thefootnote}{\fnsymbol{footnote}}

% Insert author and address list here
\begin{center}
A.~Anastassov,$^{1}$ J.~E.~Duboscq,$^{1}$ K.~K.~Gan,$^{1}$
T.~Hart,$^{1}$ K.~Honscheid,$^{1}$ H.~Kagan,$^{1}$ R.~Kass,$^{1}$
J.~Lee,$^{1}$ H.~Schwarthoff,$^{1}$ A.~Wolf,$^{1}$
% ...
\end{center} 

\small 
\begin{center}
$^{1}${Ohio State University, Columbus, Ohio 43210}\\
% ...
$^{13}${University of California, Santa Barbara, California 93106}\\
$^{14}${University of Colorado, Boulder, Colorado 80309-0390}\\
% ...
$^{25}${State University of New York at Albany, Albany, New York 12222}
\end{center}

\newpage
 
% \section{Introduction}

Although several modes of radiative and hadronic $\Upsilon$(1S) 
decays with multiparticle final states have previously been observed,
no radiative decays of the $\Upsilon$(1S) into a photon and two 
hadrons have yet been reported. Such final states have
provided the most direct evidence for two-body radiative $J/\psi$ 
decays, which are well established\cite{PDG96}
at the $10^{-3}$ level. 
To extrapolate these 
%from the $J/\psi$ 
to the
$\Upsilon$, the charge coupling to the photon and the
mass of the quark propagator predict a suppression of order 
[($q_{b}/q_{c}$)($m_{c}/m_{b}$)]$^{2}\sim$1/40.
More sophisticated calculations can be found 
in the literature\cite{Korner83}. 

The radiative decays of the $\Upsilon$(1S) can provide information on 
exotic states, including WIMP's and axions\cite{FayetKaplan91,CLEOax}.
The observation of resonances in the two-body invariant mass spectrum 
opposite photons is also one of the ways of establishing possible 
glueball candidates in radiative quarkonium 
decays, using the fact that the emission of the 
photon leaves a recoiling
``glue rich'' environment from which to form them. 
Radiative decays 
of the $\Upsilon$(1S) with charged final state hadrons have been 
studied by many experimental groups, including 
ARGUS\cite{ARG}, 
CLEO\cite{CLEOradex90}, 
and MD-1\cite{MD1}.
In the CLEO analysis, the decay modes $\Upsilon$(1S)$\,\to\!\!\gamma X$; 
$X\to\pi^+\pi^-$, $K^+K^-$ and  $p{\overline p}$ were 
investigated. 
As noted in that ref.\cite{CLEOradex90}, the only region of the 
dipion invariant 
mass distribution suggestive of an excess above background was 
in the interval 1.2-1.6~GeV,
where ten signal events were counted;
the scaled 
background in the same region corresponded to two events.
In this Letter, we extend the previous CLEO analysis, using a
new data set and exploiting many improvements 
in the performance of the CLEO~II 
detector\cite{detector},
which operates at Cornell Electron Storage Ring (CESR). 
We also present first results 
for the all-neutral final state $\Upsilon$(1S)$\,\to\gamma\pi^0\pi^0$.

% \section{Data sample and event selection}

The signal data used in this analysis were collected on the 
$\Upsilon$(1S) resonance at a center-of-mass energy 
$E_{cm} = 9.46$~GeV, corresponding to an integrated luminosity 
of 78.9 pb$^{-1}$.
Data taken at 
$E_{cm}\cong 10.52$~GeV,
just below the $\Upsilon$(4S) resonance, were used 
to subtract the $e^+e^-\!\to\gamma X$ events due to  
non-$\Upsilon$(1S) production under the resonant
$\Upsilon$(1S) peak; 
this sample 
corresponds to an integrated luminosity of 500.4 pb$^{-1}$. 
We search for events in both our $\Upsilon$(1S) (signal) 
and continuum (background) datasets compatible with the 
kinematics for the process $\Upsilon$(1S)$\,\to\gamma\pi\pi$.
Separate criteria are applied for the cases 
$\Upsilon$(1S)$\,\to\!\gamma\pi^+\pi^-$ and 
$\Upsilon$(1S)$\,\to\!\gamma\pi^0\pi^0$.

Candidate events for the $\gamma\pi^+\pi^-$ final 
state are selected as follows.
There must be exactly two 
oppositely-charged
tracks observed in the 
detector. If 
the ratio of a track's energy deposited in the calorimeter to
its momentum measured in the drift chambers is greater than 0.85,
the track is identified as an electron and the event vetoed
from further consideration. At least one of the charged 
tracks must satisfy the kinematic requirements for muon identification,
defined in terms of a track's polar angle ($\theta$) and its momentum
($p$) as $|\cos\theta|\!<$0.7 and
$p\!>$1.0~GeV/$c$. Any track satisfying these criteria and also producing
associated hits in the muon chambers is identified as a muon, and the event
is similarly vetoed. 
There must be exactly one electromagnetic shower in the good barrel 
region of the calorimeter ($|\cos\theta|\!<$0.71) with energy exceeding
0.4$\times E_{cm}$. This shower 
must have an energy deposition profile consistent with that
of a photon, and also not match, within 15 degrees, the position of
any charged track extrapolated into the calorimeter. Additional showers, 
presumed to 
be either noise or split-offs from
the tracks propagating into the 
detection volume of the calorimeter, are allowed provided their 
measured energies are each less than 500~MeV. 
The sum of the energy 
of the highest energy photon candidate plus the energies of the drift 
chamber tracks must, under the $\pi^+\pi^-$ hypothesis, lie within 
three standard deviations in energy resolution ($\sigma_E$) of 
$E_{cm}$.
Typically, we find 
$\sigma_E\cong$ 80~MeV. 
$\Upsilon$(1S)$\,\to\!\gamma K^+K^-$ and 
$\Upsilon$(1S)$\,\to\!\gamma p{\overline p}$ events, although not yet 
observed, are, if misinterpreted, more likely to fail this energy-conservation 
requirement than true $\Upsilon$(1S)$\,\to\gamma\pi^+\pi^-$ events.
The magnitude of the net momentum
vector of the event must be within three standard deviations ($\sigma_{p}$)
of zero; $\sigma_{p}$ takes into account the resolutions on the two tracks
and the high energy photon and is typically 80 MeV/$c$.
We require that the opening angle $\phi$ between
the two charged tracks satisfy the condition $\cos\phi>-$0.95. 

The momentum of each charged track recoiling against the high
energy photon is typically $\approx$ 2~GeV/$c$, beyond the momentum range 
for which the CLEO detector can cleanly separate pions from 
kaons or protons. 
We therefore
only require that the available $dE/dx$ particle 
identification information be consistent with the 
pion hypothesis. 
We have, 
nevertheless, performed dedicated searches for 
$\Upsilon$(1S)$\,\to\gamma K^+K^-$ and $\Upsilon$(1S)$\,\to\gamma 
p{\overline p}$. In neither case was any signal above background observed.

Candidate $\gamma\pi^0\pi^0$
events must have no charged tracks. The requirements on the
high-energy photon in the event are identical to the case of 
$\gamma\pi^{+}\pi^{-}$. 
Neutral pions are defined as combinations of two 
showers in the electromagnetic calorimeter with an invariant 
mass within five standard deviations
of the nominal $\pi^0$ mass. 
% At least one of the
% candidate $\pi^0$ photon daughters must be in the good barrel region
% ($|$\cos$\theta_\gamma|<$0.71); 
All photons in $\pi^0$ reconstruction
must also satisfy a minimum energy requirement ($E_\gamma$$>$50 MeV), and
have an 
energy deposition pattern consistent with true photons.
The four-momentum conservation 
requirements are also identical to the charged pion case.

We use a GEANT-based\cite{GEANT} detector simulation 
to determine the efficiency for reconstructing a radiative 
$\Upsilon$(1S) event, as a function of dipion mass, 
for each final state studied. 
Candidate $\Upsilon$(1S)$\,\to\!\gamma X$ events are generated by LUND
Monte Carlo\cite{jetset}
with flat distribution over the entire kinematically allowed
$m_X$ regime. The recoil system decays isotropically, 
and the final state pions are propagated through the
detector.
The overall event selection efficiency ($\epsilon$)
for the $\gamma\pi^+\pi^-$ final state varies smoothly from 
$\epsilon\!\cong$\,33\% at threshold ($m_{\pi^+\pi^-}\!=\!2m_\pi$)
to a maximum of $\epsilon\!\cong$\,41\% at $m_{\pi^+\pi^-}\!\cong$\,2~GeV.
By comparison, the event reconstruction efficiency for the
$\gamma\pi^0\pi^0$ final state grows rapidly from zero at threshold 
to $\epsilon\!\cong$\,30\% 
at $m_{\pi^0\pi^0}\cong$\,1~GeV and then smoothly
falls to $\epsilon\!\cong$\,28\% at $m_{\pi^0\pi^0}\!\cong$\,2~GeV.

% \section{Analysis description}

\begin{figure}[phtb]
\centerline{\epsfig{file=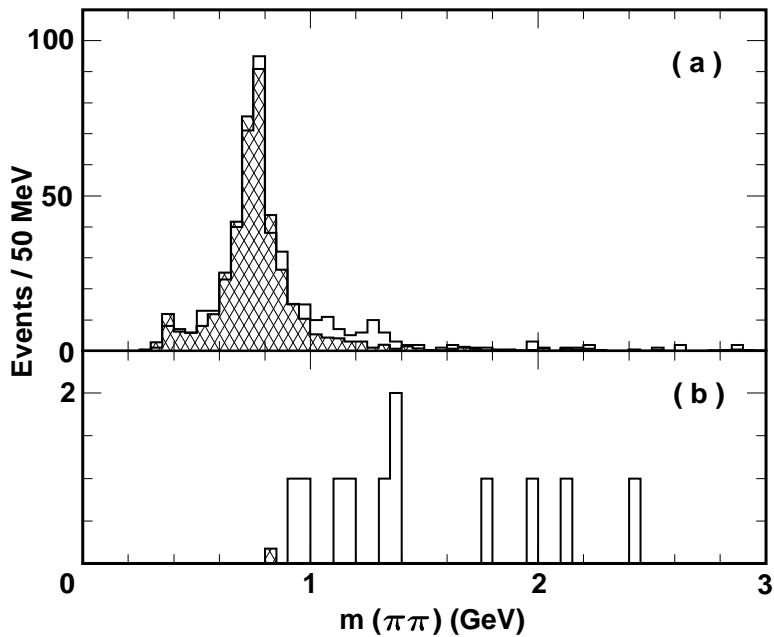,
%/cdat/tem/korolkov/1s/tex/paper-1.1.eps,
height=3.5in}}
\vspace{8pt}
\caption{ Dipion invariant mass for the $\Upsilon$(1S) data, 
with scaled continuum data (shaded) overlaid, for the final states
$\gamma\pi^+\pi^-$ (a) and $\gamma\pi^0\pi^0$ (b).} 
\label{main}
\end{figure}

The invariant mass of the recoiling hadrons 
for candidate events 
is presented in 
Fig.~\ref{main}(a) (charged pions) and 
Fig.~\ref{main}(b) (neutral pions), for both
the $\Upsilon$(1S) resonance data and the continuum data.
The continuum 
data have been properly scaled 
to the $\Upsilon$(1S) data, taking into account 
the differences in 
the luminosity of our signal and background event samples, the 
expected $1/E^{2}_{bm}$ energy dependence of the QED cross section, and
the relative event selection efficiency for the $\Upsilon$(1S) and 
the continuum data.

Prominent in Fig.~\ref{main}(a) is 
a large $\rho^0$ signal 
as verified experimentally\cite{ARG,CLEOradex90,MD1}
by other analyses. 
Backgrounds due to 
$e^{+}e^{-}\!\to\gamma\rho^0$, $\rho^0\!\to\pi^+\pi^-$ 
are expected to dominate the $\gamma\pi^+\pi^-$ analysis. Comparing the 
acceptance and luminosity-corrected signals observed in both the 
$\Upsilon$(1S) and continuum data, we note that the $\gamma\rho^0$ rate 
observed on the $\Upsilon$(1S) is consistent with the yield from the 
continuum data.
Subtracting the scaled continuum dipion mass distribution from the
resonant $\Upsilon$(1S) mass distribution in the
``$\rho$-rich region'' ($m_{\pi\pi}\!<$1.0 GeV), 
we obtain $1.4\pm 21.0$ events, consistent with zero.
Also evident 
is a small enhancement at
$m_{\pi\pi}\approx$\,400~MeV which is
largely from misidentified $e^+e^-\to\gamma\phi$; $\phi\to K^+K^-$ 
events. 
The scaled continuum data are consistent with the 
resonant $\Upsilon$(1S) yields.

We note a significant  excess of events in the $\Upsilon$(1S) data sample
over background in both the $\gamma\pi^+\pi^-$ and the 
$\gamma\pi^0\pi^0$ final states. Performing a bin-by-bin continuum
subtraction, we obtain an excess of $47.0\pm 9.3$ ($9.0\pm 3.0$) events
for the $\pi^+\pi^-$ ($\pi^0\pi^0$) data, integrated over 
$m_{\pi\pi}\!\geq$1~GeV.
We attribute these excesses to the decays 
$\Upsilon$(1S)$\,\to\!\gamma\pi^+\pi^-$ and 
$\Upsilon$(1S)$\,\to\!\gamma\pi^0\pi^0$, respectively.
Based on the total number of $\Upsilon$(1S) events in our sample
(1.86$\times 10^6$), and
correcting for the efficiencies as a function of invariant mass, we 
obtain
${\cal B}(\Upsilon$(1S)$\,\to\!\gamma\pi^+\pi^-)$=$(6.3\pm 1.2\pm 1.3)\times 
10^{-5}$ and
${\cal B}(\Upsilon$(1S)$\,\to\!\gamma\pi^0\pi^0)$=$(1.7\pm 0.6\pm 0.3)\times
10^{-5}$,
for $m_{\pi\pi}\!\geq$1~GeV, in which the 
second error is 
systematic (to be described later). 

\begin{figure}[phtb]
\centerline{\epsfig{file=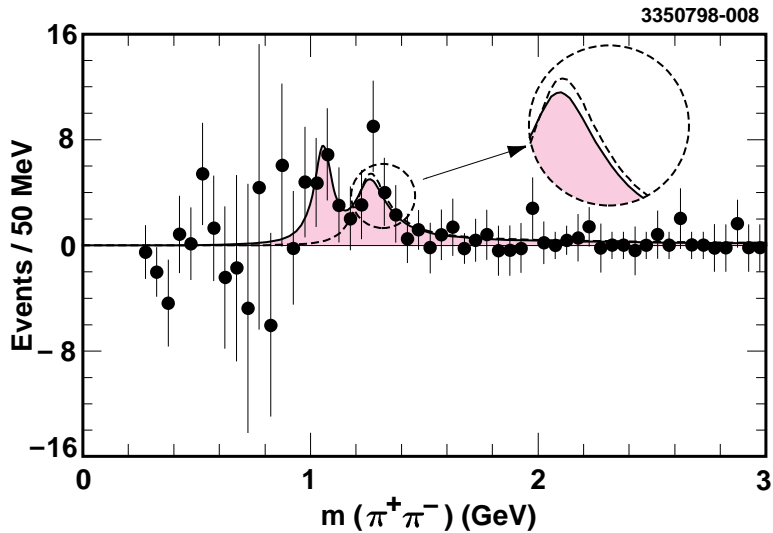,height=3.5in}}
%/cdat/tem/korolkov/1s/tex/paper-2.1.eps,
\vspace{10pt}
\caption{Fit of the continuum-subtracted
$\Upsilon$(1S) dipion invariant mass spectrum to a single $f_2$(1270)
only (dashed line, and blown up in inset) with overlay of fit to 
$f_2$(1270) plus a possible second resonance at 
$m_{\pi^+\pi^-}\!\approx$\,1.05~GeV (shaded region
under the solid line). Note that the 
$f_2$(1270) yield is relatively insensitive to the addition of
the second resonance.}
\label{two-fits}
\end{figure}

Whereas the statistics in the background-subtracted $m_{\pi^0\pi^0}$ 
mass distribution are too poor to show any obvious structure, the 
excess in the charged dipion mode is 
apparent in the region 
$m_{\pi^+\pi^-}\!\approx$\,1.0-1.4~GeV. The most prominent resonance in 
this mass range observed in radiative $J/\psi$ decays is the 
$f_2(1270)$\cite{PDG96}, for which
${\cal B}(J/\psi\!\to\!\gamma f_2(1270))$=$(1.38\pm0.14)\times 10^{-3}$.
If we assume that the excess in this interval is due to
$\Upsilon$(1S)$\,\to\!\gamma f_2(1270)$,
and neglecting any possible interference effects with other processes
producing the $\gamma\pi^+\pi^-$ final state, we can perform a fit to
the background-subtracted
on-resonance dipion invariant mass spectrum 
and thereby determine the possible level of 
$\Upsilon$(1S)$\,\to\!\gamma f_2$(1270), as shown in
Fig.~\ref{two-fits}. 
In performing this fit, we use a spin-2 Breit-Wigner signal function 
with the mean and width parameters fixed to the established\cite{PDG96} 
$f_2(1270)$ values, using the interval 0.6$-$1.8~GeV. 
Such a fit yields $34.8 \pm 9.7$ events 
with a $\chi^2$ per 
degree of freedom of 12.8/23. If, instead, we allow the mass and width to
float,
we obtain a yield of $30.1^{+9.9}_{-9.3}$ events, with a fitted mass of
$(1.28\pm0.02)$~GeV and a width of $(100^{+80}_{-40})$~MeV.
Assuming no other contributions to the spectrum in
Fig.~\ref{two-fits},
the corresponding efficiency-corrected 
product of branching fractions would be
${\cal B}(\Upsilon$(1S)$\,\to\gamma f_2(1270))\times 
{\cal B}(f_2(1270)\to\pi^+\pi^-) =
(4.6\pm 1.3^{+1.6}_{-1.5})\times 10^{-5}$, which gives\cite{PDG96} 
${\cal B}(\Upsilon$(1S)$\,\to\!\gamma f_2(1270))=
(8.1\pm 2.3^{+2.9}_{-2.7})\times 10^{-5}$. 
The likelihood that the excess in this region is due to an upward fluctuation
of background is determined to be less than 0.01\%.
%Note: A simple look at having the background fluctuate in
%a Poisson fashion to the observed 4-bin distribution has a 
%probability of approx. 3 10^{-7}.

If this excess is due to 
$\Upsilon$(1S)$\,\to\!\gamma f_2(1270)$, 
then, by isospin, we expect to also observe 
$\Upsilon$(1S)$\,\to\!\gamma f_2(1270)$, $f_2$(1270)$\,\to\!\pi^0\pi^0$ 
at half the charged rate.
When compared to the fitted result in the charged dipion case, the 
excess of
$10.8 \pm 3.3$ events observed is consistent, after 
efficiency-correction, with this expectation. 
%If we use the same fit parameters for the $f_2$(1270) and assume
%no other contributions to
%that the $f_2$(1270) saturates 
%Fig.~\ref{main}(b), we obtain $10.8\pm 3.3$ events,
%for an isospin ratio of $0.53^{+0.33+0.32}_{-0.19-0.15}$.
%consistent (after efficiency-correction) with the ratio expected from isospin.
The probability of the continuum background fluctuating up to 
the $\pi^0\pi^0$ signal is negligible. Non-$f_2$ contributions, 
if any, are difficult to assess with the limited statistics of the 
signal sample.

Although there is no resonance with $m \approx 1.05$~GeV  
expected in our sample,
we nevertheless note an apparent 
enhancement in this mass region in Fig.~\ref{two-fits}. 
To investigate this further, we have allowed for a second Breit-Wigner in 
our fit, with the values of mass and width for this 
second Breit-Wigner allowed to float, but with the $f_2$ parameters again 
constrained to the established values (Fig.~\ref{two-fits}). 
We then obtain $20.5\pm 12.3$ events for this second 
Breit-Wigner, at a mass of $(1.05\pm 0.02)$~GeV and a fitted width of 
$(100\pm 90)$~MeV; the putative $f_2$(1270) yield correspondingly
drops to $29.7\pm 11.0$ events. The overall $\chi^2$ per degree of freedom 
in this second fit improves to 6.2/20. We note that 
although the 
level of the excess at 1.05~GeV is at the 1.7 standard deviation level, 
the likelihood that the excess in the 1.28~GeV mass region is an
upward fluctuation of background
is still small (less than 0.1\%).

We have considered possible contamination to our 
signal from the 
process $\Upsilon$(1S)$\,\to\!\gamma\mu^+\mu^-$.
This is
evaluated by selecting, rather than vetoing, 
events having a high energy photon and two charged 
tracks, in which there are hits in the muon chambers matched to 
at least one of
the charged tracks. For such a search, the continuum-subtracted
$\Upsilon$(1S) data yields $18\pm 19$
$\Upsilon$(1S)$\,\to\!\gamma\mu^+\mu^-$ event
candidates. 
Knowing that the maximal inefficiency for $\gamma\mu^+\mu^-$ events is
30\%,
%Knowing the probability that a muon will pass our veto requirement,
we determine that the contribution from the $\gamma\mu^+\mu^-$ final state
to our signal
$\Upsilon$(1S)$\,\to\!\gamma\pi^+\pi^-$ sample 
has a central value less than 
5.4 events, and consistent with zero;
we include this in our systematic uncertainty.

The decay $\Upsilon$(1S)$\,\to\!\pi^0\pi^+\pi^-$, 
although not yet 
observed, could produce background to the $\gamma\pi^+\pi^-$ final state 
in the cases in which the $\pi^0$ decays either very asymmetrically
(resulting in one very high energy shower with energy almost
equal to the energy of the parent $\pi^0$), or
produces a $\pi^0$ in which both decay photons are approximately collinear,
and cannot be distinguished. 
In the latter case, the photon showers overlap and merge into a single 
detected calorimeter shower. 
We have
conducted a dedicated search for the 
decay $\Upsilon$(1S)$\,\to\!\pi^0\pi^+\pi^-$, 
with identical requirements on the $\pi^{0}$ candidate,
on the tracks of the two charged pion candidates,
and on four-momentum conservations as in the $\gamma\pi\pi$ analysis.
%very similar to our 
%$\Upsilon$(1S)$\,\to\!\gamma\pi^+\pi^-$ search.  
%We form $\pi^0$ candidates as before from photon pairs; charged pions 
%must satisfy the same track requirements imposed for the 
%$\Upsilon$(1S)$\,\to\!\gamma\pi^+\pi^-$ measurement. Overall
%event momentum and energy-conservation is also required, as before, at the
%three standard deviation level.
This search was consistent with zero events being found.
The associated upper limit is
 ${\cal B}(\Upsilon({\rm 1S})\to\pi^0\pi^+\pi^-)<1.84\times 10^{-5}$ at
90\% confidence level. 
Based on this null result and 
the probability to 
misinterpret a ``merged'' $\pi^0$ 
as a photon, we would expect fewer than 3.4 events 
contamination from the decay $\Upsilon$(1S)$\,\to\!\pi^0\pi^+\pi^-$ 
in our $\gamma\pi^+\pi^-$ event sample over the entire kinematically 
allowed dipion invariant mass range. The net contribution from 
$\Upsilon$(1S)$\,\to\!\pi^0\pi^+\pi^-$ events in which the
$\pi^0$ decays asymmetrically is determined to be less than 0.7 events.

We note that  
two pions produced in
$\Upsilon$(1S)$\,\to\!\gamma f_2$(1270), 
$f_2$(1270)$\,\to\!\pi^+\pi^-$ will have
a characteristic
angular distribution, due to the tensor nature of the $f_2$(1270). 
We have correspondingly fit the 
helicity angle distribution (defined as the angle between one of the pions
and the dipion parent measured in the dipion rest frame)
for the mass interval
1.2-1.4~GeV, after subtracting out the contribution from 
the $\gamma\rho$ final
state. Such a fit gives confidence levels of 48\%, 35\%, 0.0\%, and 
0.1\% for the system recoiling against the photon to be
a tensor, scalar, vector or axial vector, respectively.
Although inconclusive on its own,
this spin-parity
analysis of the dipion system 
strongly
favors a tensor or scalar
assignment for the dipion system, and rules out a
vector or axial vector interpretation.

For the measurement $\Upsilon$(1S)$\,\to\!\gamma\pi^+\pi^-$, systematic 
uncertainties are due primarily to the muon veto used to suppress the 
$\gamma\mu^+\mu-$ final state (12\% relative error), uncertainties in our 
total efficiency (5\%, arising mainly from event
triggering uncertainties), and our uncertainty in the total
number of $\Upsilon$(1S) events (3\%). Because 
we have assumed that the photon angular distribution is isotropic
in our Monte Carlo event generator, there is an additional uncertainty
%systematic error 
(16\%) from our extrapolation to the region
$|\cos\theta_\gamma|\!>$0.71.
For the possible 
$\Upsilon$(1S)$\,\to\!\gamma f_2$(1270) signal, we have an additional
systematic error (20\%) due to the fitting procedure used to extract
the signal, including the
possible effect of the apparent enhancement in the 
region $m_{\pi^+\pi^-}\!\approx$\,1.05~GeV, and asymmetric 
uncertainties due to the possible interference between events from
the $\Upsilon$(1S)$\,\to\!\gamma f_2$(1270) excess and $\pi^+\pi^-$ pairs not 
associated with either the $\rho^0$ or the resonant enhancement 
($^{+20}_{-15}$)\%.

For the $\gamma\pi^0\pi^0$ final state, primary uncertainties in our 
integrated measurement ${\cal B}(\Upsilon$(1S)$\,\to\!\gamma\pi^0\pi^0)$ 
are due to the possible anisotropy of $\Upsilon$(1S)$\,\to\!\gamma\pi^0\pi^0$
decay (16\%), $\pi^0$ finding (8\%), trigger efficiency
(4\%), and the number of $\Upsilon$(1S) events (3\%). 

In summary, we have made the first observation of the radiative
decay $\Upsilon$(1S)$\,\to\!\gamma\pi\pi$ in both charged and neutral modes.
Restricted to 
%the high dipion mass regime 
$m_{\pi\pi}\!\geq\,$1.0 GeV, we obtain
${\cal B}(\Upsilon$(1S)$\,\to\!\gamma\pi^+\pi^-$)=$(6.3\pm 1.2\pm 1.3)\times 
10^{-5}$, and
${\cal B}(\Upsilon$(1S)$\,\to\!\gamma\pi^0\pi^0$)=$(1.7\pm 0.6\pm 0.3)\times 
10^{-5}$.

The $\pi^+\pi^-$ mass and helicity angle distributions are suggestive
of $f_2$(1270) production as a source.
Under the 
$\Upsilon(1S)\,\to\!\gamma f_2(1270)$ assumption, the efficiency-corrected 
product of branching fractions of this enhancement corresponds to 
${\cal B}(\Upsilon(1S)\,\to\!\gamma f_2(1270))\times 
{\cal B}(f_2(1270)\,\to\!\pi^+\pi^-) =
(4.6\pm 1.3^{+1.6}_{-1.5})\times 10^{-5}$. 
% or ${\cal B}(\Upsilon$(1S)$\,\to\!\gamma f_2(1270))=
% (8.1\pm 2.3^{+2.8}_{-2.6})\times 10^{-5}$. 
In the $\pi^0\pi^0$ mode, the net yield relative to the 
charged mode and the shape of the 
$\pi^0\pi^0$ mass spectrum are also consistent with 
$\Upsilon$(1S)$\to\gamma f_2$(1270). 
This is approximately 
twice the rate that would be expected by the 
[($q_{b}/q_{c}$)($m_{c}/m_{b}$)]$^{2}$
scaling from 
${\cal B}(J/\psi\to\gamma f_2(1270))$.

\vspace{0.25cm}
\centerline{\bf Acknowledgments}
\vspace{0.25cm}

We gratefully acknowledge the effort of the CESR staff in providing us with
excellent luminosity and running conditions.
This work was supported by 
the National Science Foundation,
the U.S. Department of Energy,
Research Corporation,
the Natural Sciences and Engineering Research Council of Canada, 
the A.P. Sloan Foundation, 
the Swiss National Science Foundation, 
and the Alexander von Humboldt Stiftung.

\end{document}